\documentclass[12pt]{iopart}

\usepackage{graphicx,graphics,times,color}
\usepackage{bm}
\usepackage{longtable}
\usepackage{color}
\usepackage{graphicx}
\usepackage{epstopdf}

\def\be{\begin{equation}}
\def\ee{\end{equation}}
\def\ba{\begin{eqnarray}}
\def\ea{\end{eqnarray}}
\def\la{\langle}
\def\ra{\rangle}

\def\Htot{H_\mathrm{tot}}

\begin{document}

\title[Quantum dot cellular automata]
{Quantum dot spin cellular automata for realizing a quantum processor}

\author{Abolfazl~Bayat$^1$, Charles~E~Creffield $^2$,
John~H~Jefferson$^3$, Michael~Pepper$^4$ and Sougato~Bose$^1$}

\address{$^1$ Department of Physics and Astronomy, University College London,
Gower Street, London WC1E 6BT, United Kingdom}

\address{$^2$ Departamento de F\'isica de Materiales, Universidad
Complutense de Madrid, E-28040, Madrid, Spain}

\address{$^3$ Department of Physics, Lancaster University, Lancaster LA14YB, United Kingdom}

\address{$^4$ Department of Electronic and Electrical Engineering, University College London,
Gower Street, London WC1E 6BT, United Kingdom}

\ead{c.creffield@fis.ucm.es}

\vspace{10pt}
\begin{indented}
\item[]\today
\end{indented}

\begin{abstract}
We show how single quantum dots, each hosting a singlet-triplet qubit, can be placed
in arrays to build a spin quantum cellular automaton.
A fast ($\sim10$ ns) deterministic coherent singlet-triplet filtering, as opposed to current incoherent tunneling/slow-adiabatic based quantum gates (operation time $\sim300$ ns), can be employed
to produce a two-qubit gate through capacitive (electrostatic) couplings that can operate over significant distances. This is the coherent version of the widely discussed charge and nano-magnet cellular automata, and would increase speed, reduce dissipation, and perform quantum computation while interfacing smoothly with its classical counterpart. This combines the best of two worlds -- the coherence of spin pairs known from quantum technologies, and the strength
and range of electrostatic couplings from the charge-based classical cellular automata. Significantly our system has
zero electric dipole moment during the whole operation process,
thereby increasing its charge dephasing time.
\end{abstract}


\maketitle

\section{Introduction}

A coherent version of the widely discussed charge and nano-magnet cellular automata \cite{lent-porod,Cowburn,Lent} would offer increased speed, 
reduced dissipation, and would perform quantum computation while interfacing smoothly with its classical counterpart. However, maintaining long time coherence is a challenge \cite{Cellular-automata}. It is appealing to use
quantum dot (QD) spins, with coherence times of $\sim260$ $\mu$s, \cite{Loss-DiVincenzo,Yacoby-coherence,Taylor-coherence,Hanson-Burkard,petta,marcus-CZ} and
in particular singlet-triplet electron pairs, which are largely decoherence free \cite{petta,marcus-CZ,yacobi-2QBgate}. Here we show how ``single" QDs, each hosting a singlet-triplet qubit, can be placed
in arrays to build a spin quantum cellular automaton.
Our proposal combines the best of two worlds -- the coherence of spin pairs known from quantum technologies, and the strength
and range of electrostatic couplings from charge-based classical cellular automata.
A fast ($\sim 10$ ns) deterministic two-qubit gate is accomplished via non-equilibrium (non-adiabatic) dynamics and capacitive interactions in the course of which no electric dipole moment ever arises, thereby increasing charge coherence significantly.

Many double dot proposals for two-qubit gates already exist, using electrostatic interactions \cite{Hanson-Burkard,petta,marcus-CZ,yacobi-2QBgate,Taylor}. Their nonzero dipole moment during the gate operations, however, causes rapid charge dephasing \cite{marcus-CZ}. Using more symmetric charge configurations, on the other hand, makes the gate operation slow ($\sim 150$ ns) \cite{yacobi-2QBgate}. If charge tunneling in double QDs eventually becomes incoherent due to the long time scale and strong dephasing, then a set time for the gate operation will disappear, rendering the system indeterministic. In view of the increasing speed of control electronics it is thereby worthwhile to consider non-adiabatic tunnelings that do not create dipole moments. Our proposal can also be realized in a ring of four coupled QDs, which is functionally equivalent to the system we study.
This ring structure has already been used for charge-based qubits \cite{Statce-4QD-charge-qubit}, but the spin dependent dynamics has not yet been explored.

Singlet-triplet qubits in double dots face a fundamental obstacle by seeking to exploit coherent charge tunneling for two-qubit gate operations. Indeed, the very limited and short charge dephasing time ($\sim$ 1 ns) is comparable with the charge tunneling times. For instance the 6 $\mu eV$ tunneling rate in Ref.~\cite{marcus-CZ} gives a period of $0.7$ ns for coherent oscillations. According to Ref.~\cite{Loss-dipole}, the dominant source of dephasing in double dot systems is the interaction between the electric dipole of the two electrons with random electric field fluctuations. In double dot systems the asymmetric charge configuration $(0,2)$
has a large dipole moment $\overrightarrow{\mathbf{p}} \sim e \overrightarrow{\mathbf{d}}$, where
$\overrightarrow{\mathbf{d}}$ is the separation between the dots, which gives rise to
charge decoherence. Motivated by this, we present a system which benefits from an extra charge orbital that always keeps the charge distribution symmetric with zero dipole moment, resulting in much longer charge dephasing times. One can call this a singlet-triplet qubit, which couples to its environment (and indeed any distant qubits in a scalable network with which the couplings are not sought) through its {\em quadrupole} moment, as opposed to the dipole.
In addition this extra orbital allows for symmetric charge configurations during gate operations, which enables the simultaneous implementation of identical two-qubit gates between all neighboring pairs in a row, as required for generating cluster states for measurement-based quantum computation. Moreover, our quench dynamics is applicable for degenerate qubit (i.e. singlet-triplet) levels so that no relative phase develops between them during storage (non-operative) periods of the qubit. Both the above features are absent in double dot singlet-triplet qubits, because of their asymmetric charge configurations and the need for an initial singlet-triplet gap for adiabatic operation at non-zero speeds.

\begin{figure}
\centering
\includegraphics[width=.52\textwidth,clip=true]{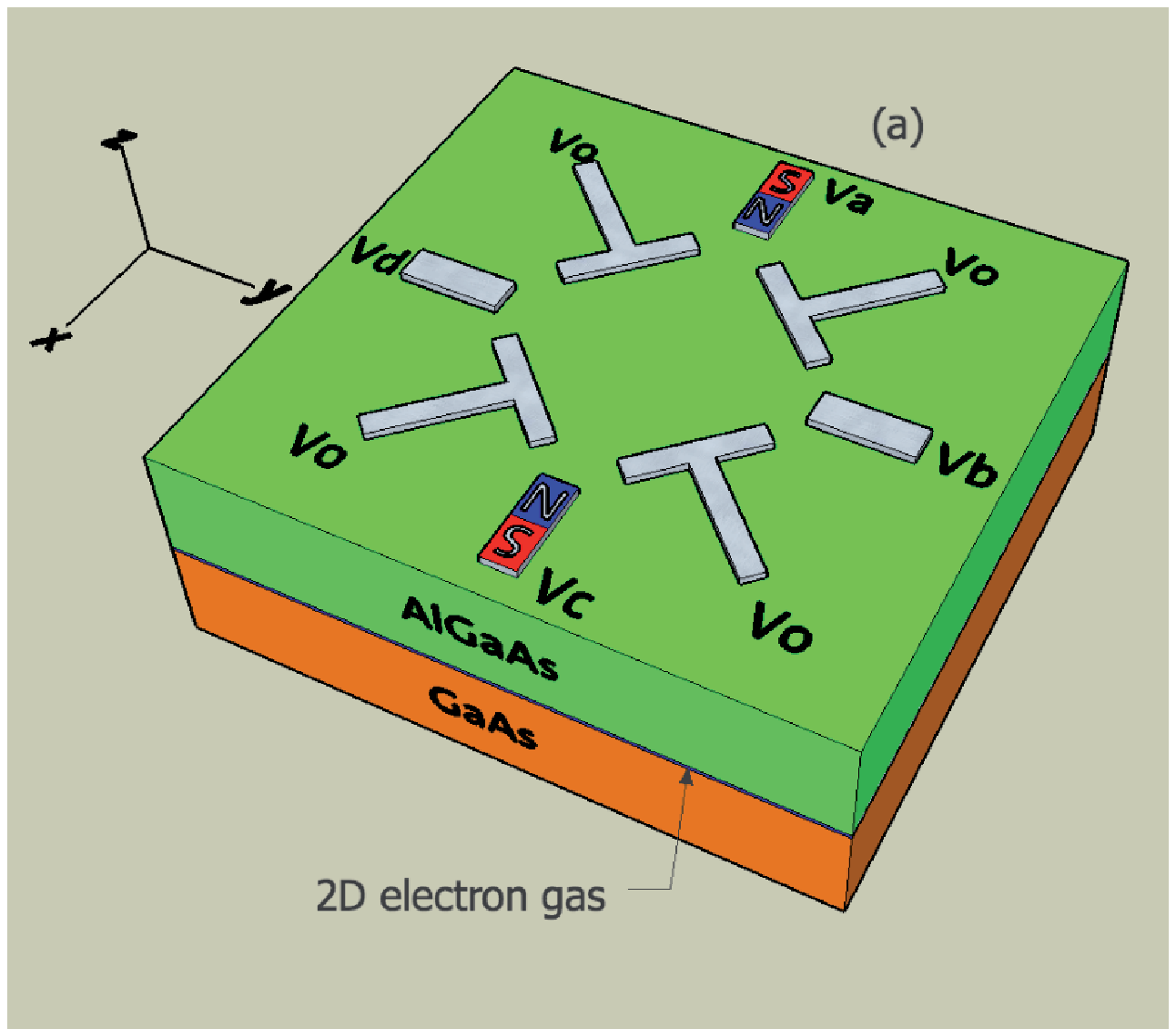}
\includegraphics[width=.35\textwidth,clip=true]{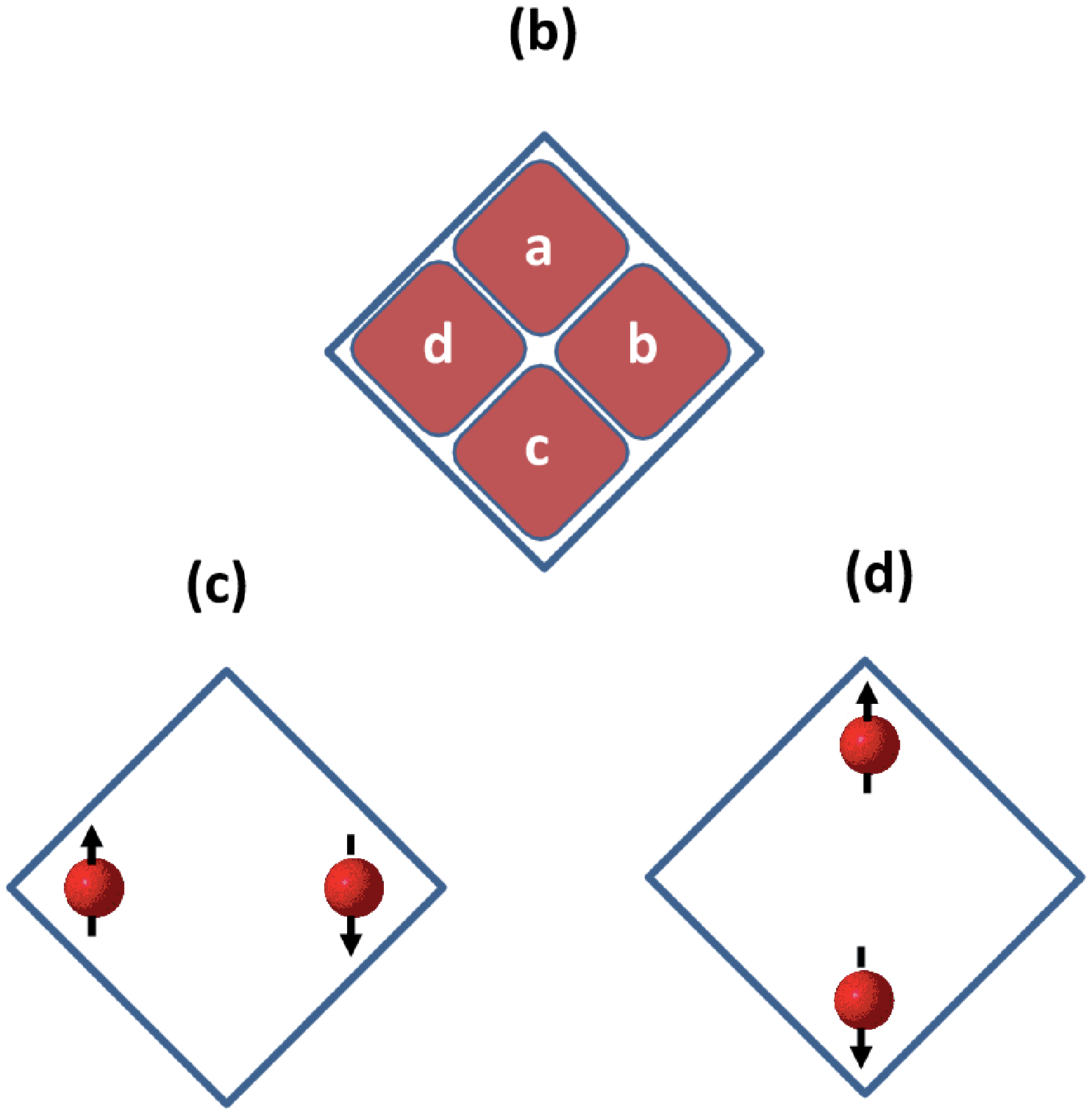}
\caption{(a) Three dimensional cross-section of the heterostructure junction,
showing the gate configuration and location of micro-magnets.
As indicated, the $\hat x$ and $\hat y$ directions lie in the plane of
the 2D electron gas, while $\hat z$ is perpendicular to it (i.e. the growth direction of the heterostructure).
A large negative voltage $V_0$ is applied to the ``T'' gates to impose
high potential barriers on the 2D electron gas lying between the
AlGaAs and GaAs layers. These barriers define the edges of the square QD.
The four remaining finger gates ($V_a$, $V_b$, $V_c$, and $V_d$)
are used to adjust the local potential
in each quadrant of the QD, shown in (b), and thus to control
the localization of the two electrons. For simplicity we assume
the local potentials are constant within each quadrant.
Gates $V_a$ and $V_c$ are also micro-magnets, used to 
rotate the spins of the electrons (see Section \ref{single_qubit}).
(b) Labeling of the four quadrants of the square QD.
The potential in quadrant $a$ is controlled by gate $V_a$, that of
quadrant $b$ by $V_b$, and so on. Each quadrant has approximate dimensions
of $L /2 \times L / 2$, where $L$ is the side-length of the QD.
(c) Charge configuration of $|S_\leftrightarrow\ra$ and $|T_\leftrightarrow^0\ra$,
in which the electron density is strongly peaked in quadrants $b$ and $d$.
(d) Charge configuration of $|S_\updownarrow\ra$ and $|T_\updownarrow^0\ra$; in contrast to the previous case the electron density is now localized
in quadrants $a$ and $c$. The notation used in the paper of $\updownarrow$ 
and $\leftrightarrow$ indicate the $x$ and $y$ directions respectively in 
these diagrams.}
\label{fig1}
\end{figure}

\section{Two electrons in a square quantum dot}

We consider a system of two electrons held in a square semiconductor QD with a hard-wall boundary, approximately realizable by gating a two-dimensional
electron gas at a heterojunction interface as shown in \Fref{fig1}(a).
To describe this system we take as our starting point the effective-mass Hamiltonian for the two interacting electrons:
\begin{equation}\label{H_schrodinger}
H = -\frac{\hbar^{2}}{{2m^{\ast}}}\left[\nabla_{1}^{2}+\nabla_{2}^{2}\right]+\frac{{e^{2}}}{{4\pi\varepsilon|{\bf {r}}_{1}-{\bf {r}}_{2}|}}
+ V_c({\bf {r}}_{1})+V_c({\bf {r}}_{2})
+ V_{g}({\bf {r}}_{1})+V_{g}({\bf {r}}_{2}) , \quad
\end{equation}
where $V_c(\mathbf{r})$ is the confinement potential and $V_g(\mathbf{r})$ is the potential energy due to external gates. 
The cross-sectional schematic of the heterostructure interface and the gate configuration of the square quantum dot is shown in \Fref{fig1}(a).
For simplicity we divide the square QD into four quadrants, as shown in \Fref{fig1}(b),
and apply a constant potential to gates $b$ and $d$ giving an electron potential energy $V_g=V$ in quadrants $b$ and $d$, while $V_g=0$ when the electrons are in quadrants $a$ and $c$ . When $V$ is positive the electron density is enhanced in quadrants $a$ and $c$ (and depleted in quadrants $b$ and $d$),
and vice versa when $V$ is negative.

\begin{figure}
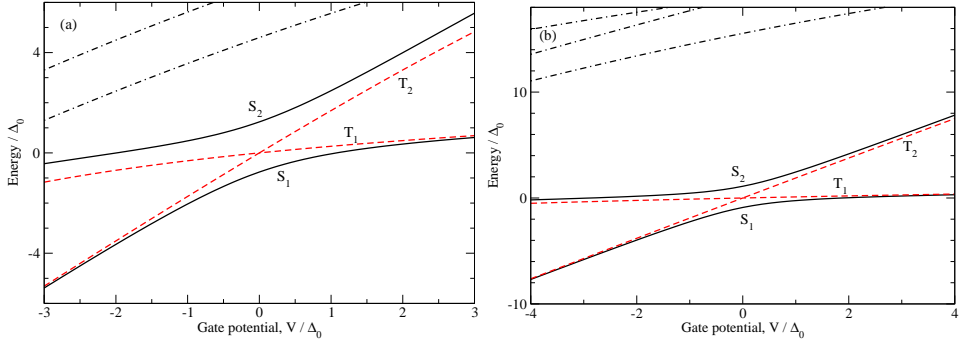

\centering
\includegraphics[width=0.4\textwidth,clip=true]{fig2a}
\includegraphics[width=0.4\textwidth,clip=true]{fig2b}
\caption{Lowest energy levels of a GaAs QD, obtained by the
exact diagonalization of \Eref{H_schrodinger}. 
(a) For a QD of size $L = 400$nm.
Singlet states are
shown with solid (black) lines, triplet states with dashed (red) lines.
The lowest multiplet consists of the singlet ground state $| S_1 \ra$
and excited state $| S_2 \ra$, and two $S_z = 0$ triplet states $| T_1 \ra$ and
$| T_2 \ra$. At $V=0$, the two triplet states are degenerate, and the singlet
states have an energy splitting of $2 \Delta_0$, which we use as the
unit of energy. In this case the energy splitting has the
value $\Delta_0 \simeq 20 \mu$eV. The black dash-dotted lines
indicate the next highest energy levels; near $V=0$ the lowest
multiplet of four levels is well-isolated from the rest of the spectrum.
(b) As in (a) but for a QD of size $L = 800$ nm.
The lowest multiplet of states has the same form as in (a), and is even more
isolated from the next highest states. The energy splitting reduces
as $L$ increases, however, and for this QD size $\Delta_0 = 2.1 \mu$eV. 
This reduction in the energy scale would require lower operating
temperatures, and would also give slower qubit operation times.} 
\label{fig2}
\end{figure}

The time-independent Schr\"odinger equation for the Hamiltonian \eref{H_schrodinger} may be solved numerically. Since total spin is a good quantum number the eigenstates are singlets and triplets. Furthermore, as the total wave function factors into the product of a spatial part and spin part, we need only solve for the spatial component. Under interchange of electron coordinates 
this will be symmetric for singlet states, and antisymmetric for triplets.
In \Fref{fig2} we show the lowest-lying energy levels for two sizes
of QD, $L = 400$nm and $L = 800$ nm, as a function of the gating potential $V$. 
Throughout this work we shall use material parameters for GaAs, and
as the effective Bohr radius for electrons in this material
is $a_B \simeq 8.8$nm, these two QD sizes correspond to $L=45 a_B$ and 
$L = 90 a_B$ respectively. We see that in both cases the energy level
structure is similar for small $V$, consisting of a multiplet of two 
singlets and triplets, well separated from the next higher states.
The formation of this isolated multiplet is a general feature of large QDs 
for which $L \gg a_B$. When this condition is satisfied
the Coulomb interaction dominates the kinetic energy, 
causing the electronic charge density to localize near the corners of 
the QD \cite{john1}.

We emphasise that although our model and gating scheme are rather simple,
our approach does not require perfect square symmetry, or hard walls,
or a perfectly flat background potential. In experiment, for example,
gating will produce soft-wall
confinement, but this effect, together with deviations from
symmetry, and roughness in the confining potential arising from disorder,
can be accounted for by renormalizing the tunneling (see \Eref{Heff})
between the two charge configurations. The formation of these low-lying
localized states is a rather robust effect.

We will use the states in the lowest multiplet as our qubit space.
At $V=0$ the two triplet states are degenerate, and the two
singlet states have an energy splitting of $2 \Delta_0$. 
To choose an optimum size for the QD we must balance two
opposing effects. As we can see from \Fref{fig2}, in larger
QDs the ground state multiplet in the larger
QD is more isolated from higher states than for smaller QDs,
which will reduce leakage from the qubit space. 
The tradeoff from using large QDs, however, is that
the energy splitting $\Delta_0$ drops rapidly with $L$ \cite{john1}, making the
qubit operation time longer and putting more stringent limits on the
operating temperature of the device (see Sec. \ref{practicality}).
We therefore use the value of $L = 400$nm in our study, which provides
a reasonable compromise between these effects.

\begin{figure*}
\centering
{
\includegraphics[width=.22\textwidth,angle=0]{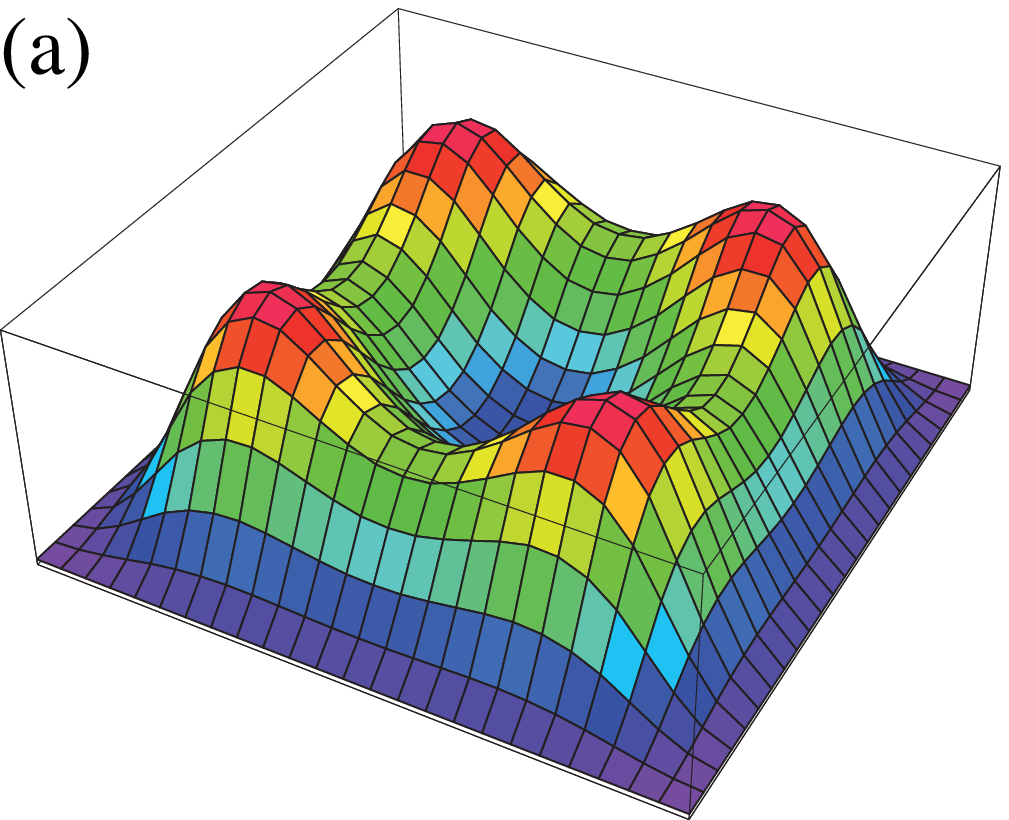}
\includegraphics[width=.22\textwidth,angle=0]{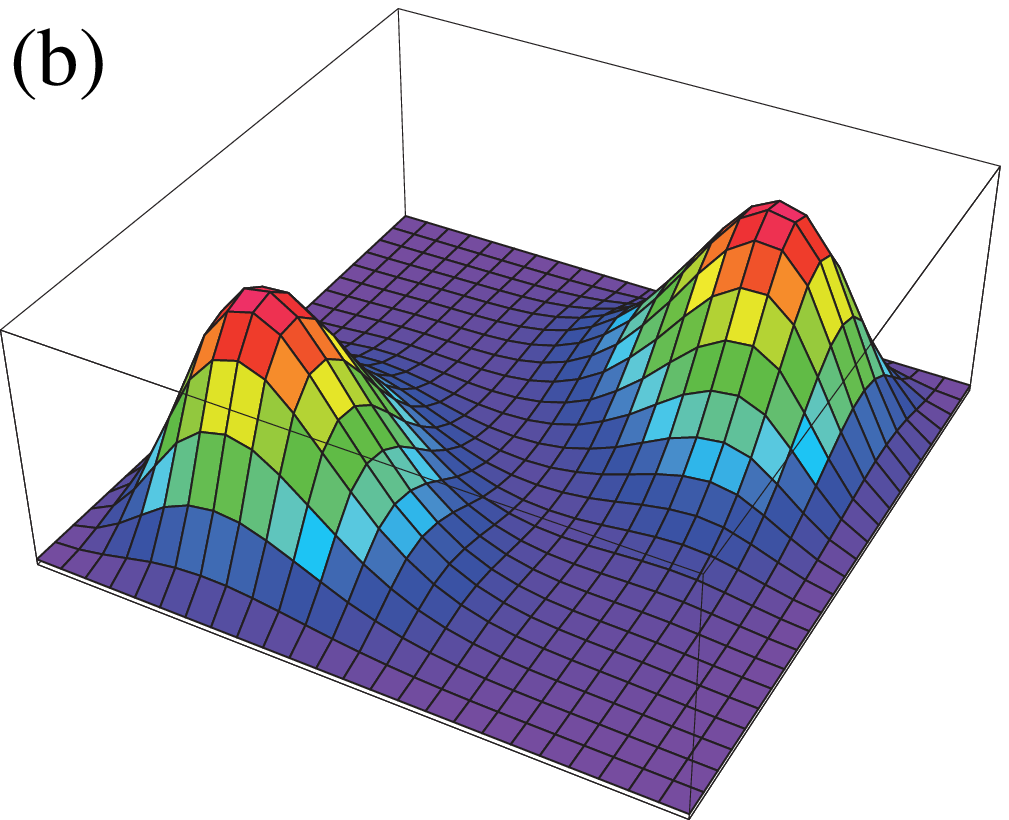}
\includegraphics[width=.22\textwidth,angle=0]{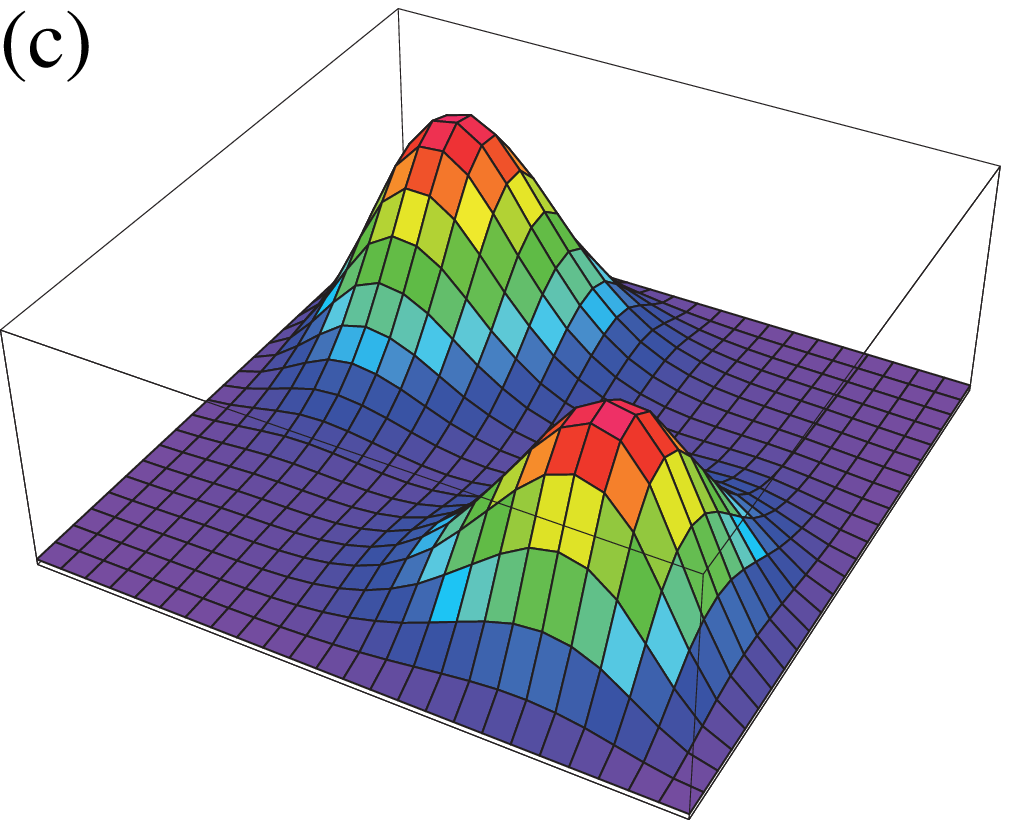}
\includegraphics[width=.22\textwidth,angle=0]{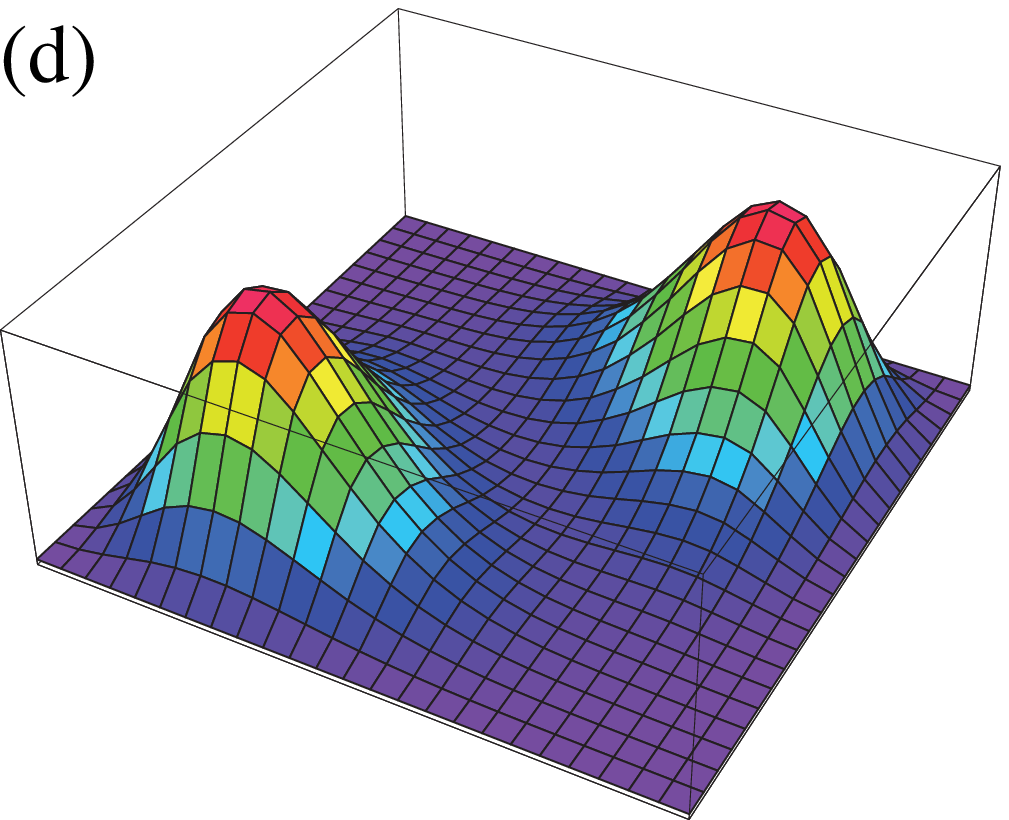}}
\centering
{
\includegraphics[width=.15\textwidth,angle=45]{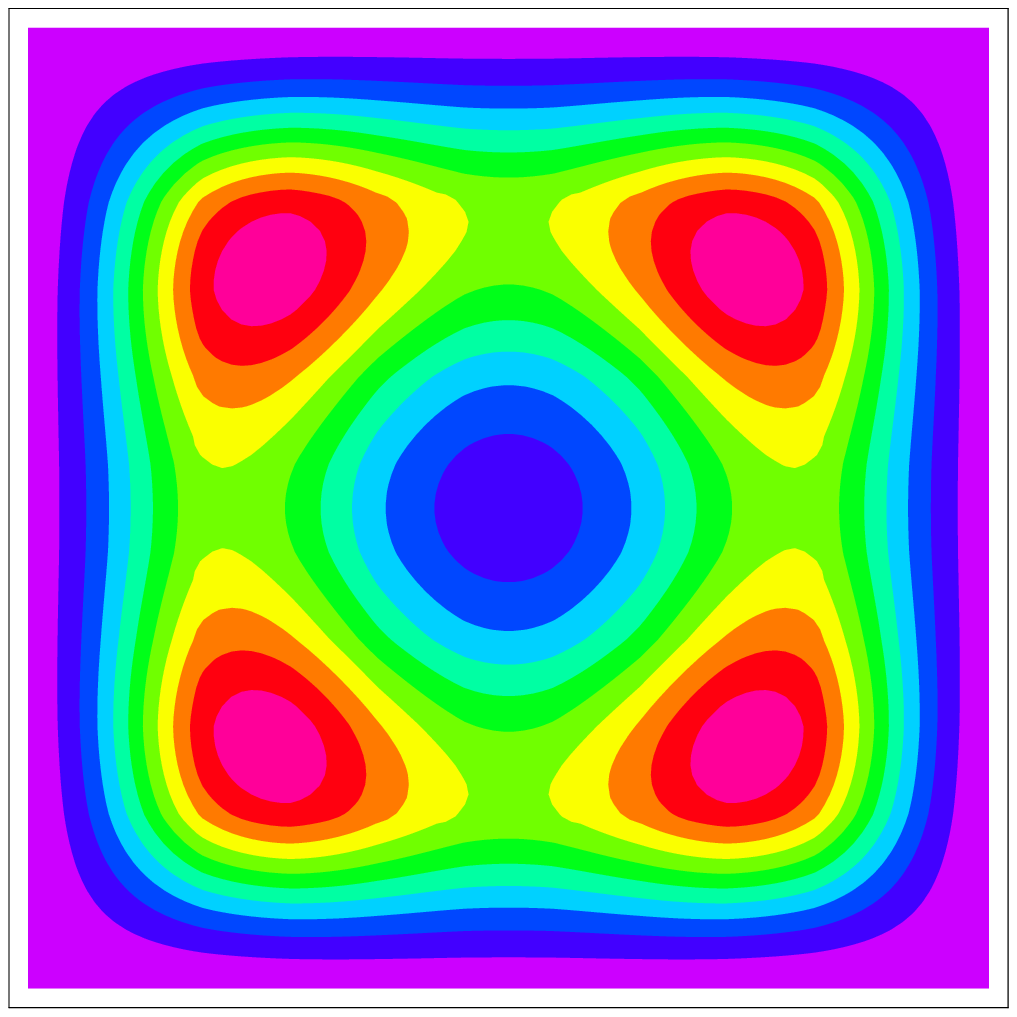}
\includegraphics[width=.15\textwidth,angle=45]{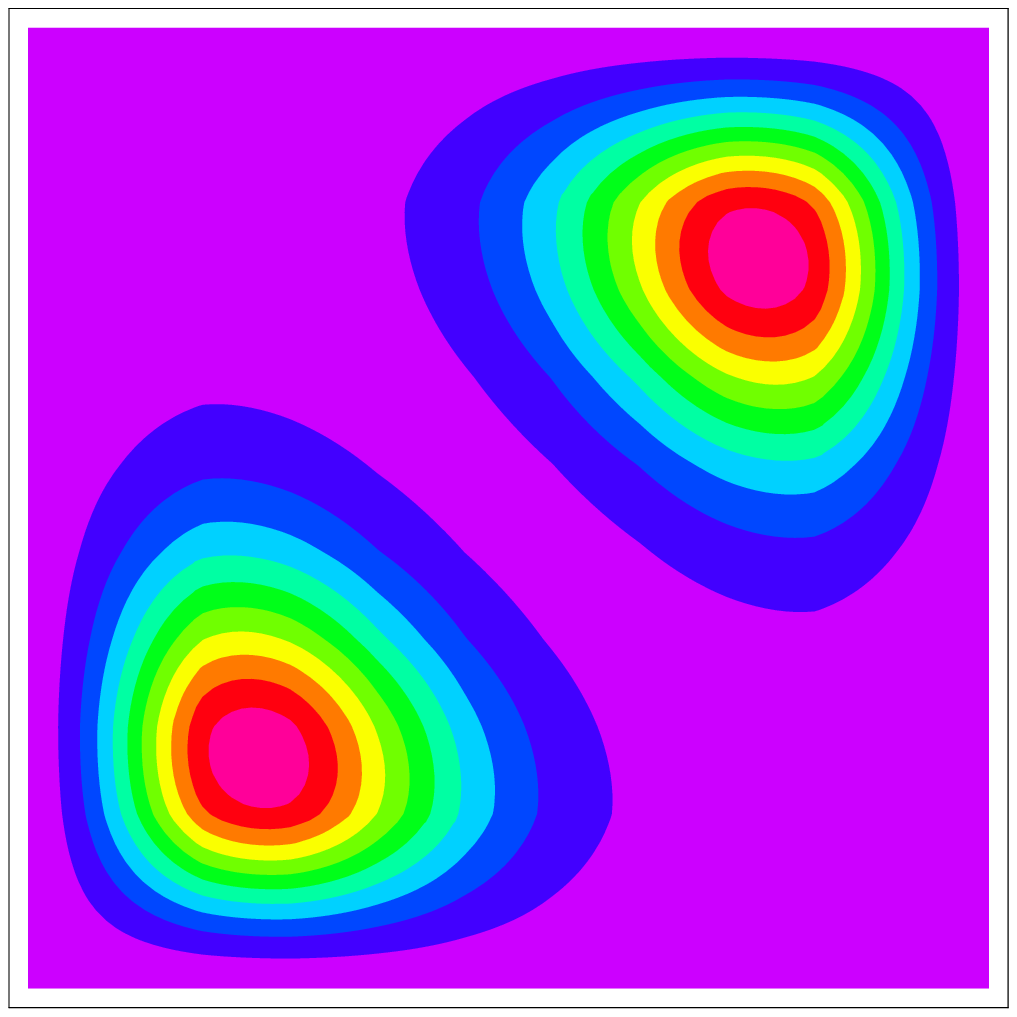}
\includegraphics[width=.15\textwidth,angle=45]{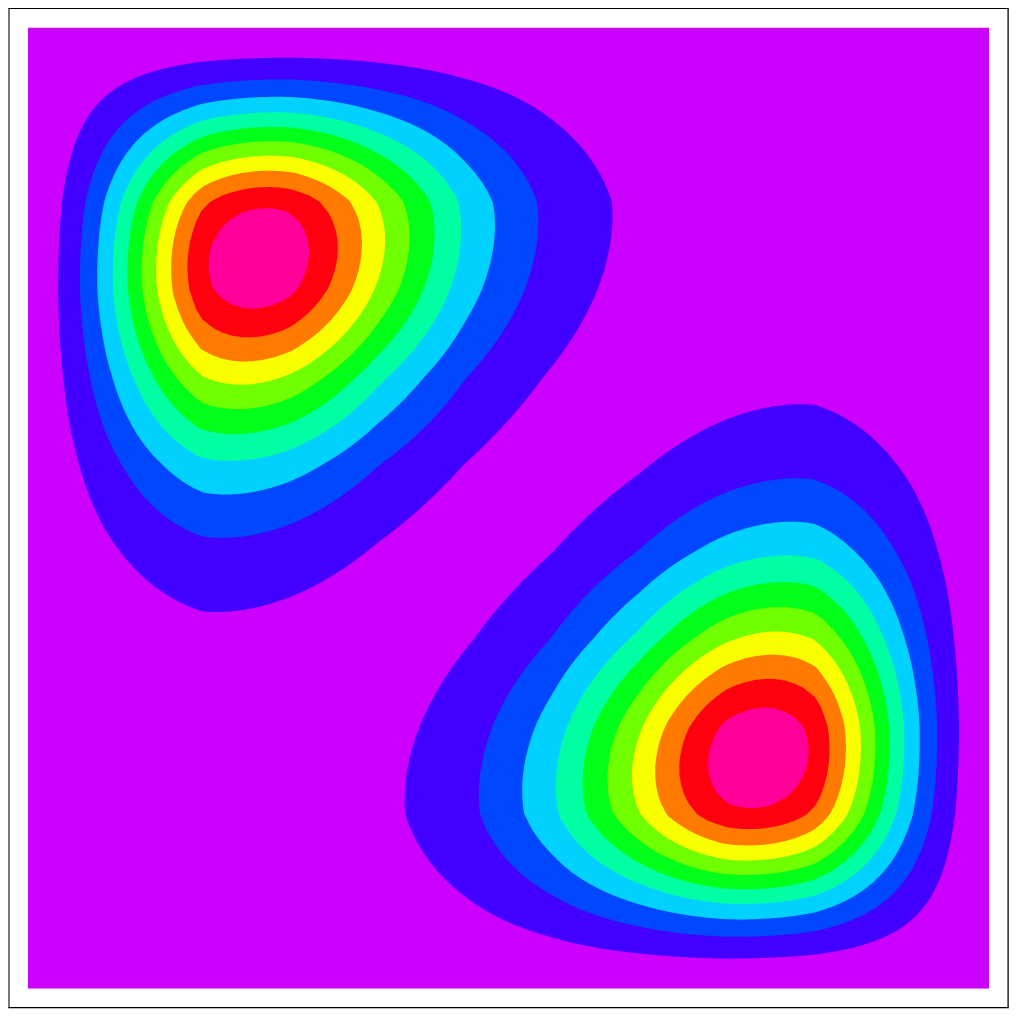}
\includegraphics[width=.15\textwidth,angle=45]{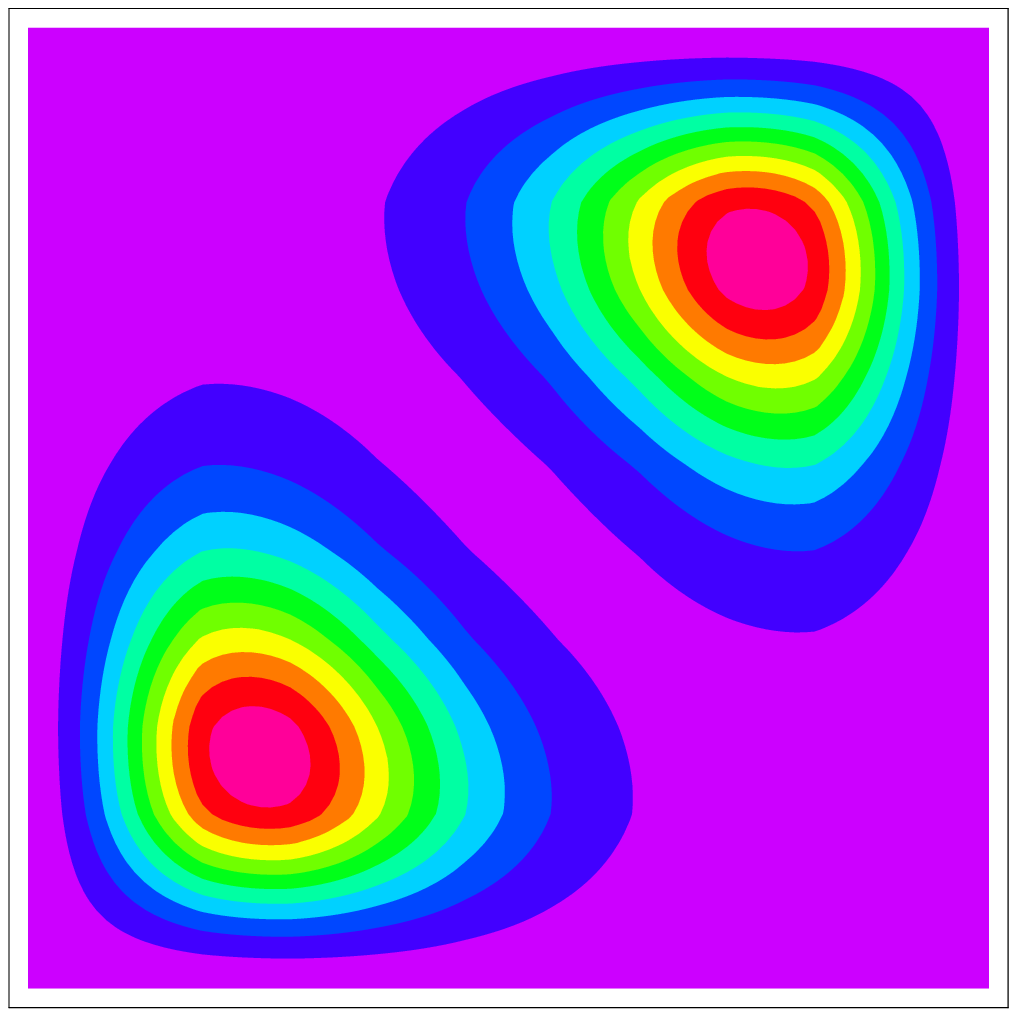}}
\caption{The charge distribution of the two electrons and their contour representations in a square QD of size $L=400$ nm.
(a) The charge distribution of $|S_1(V=0)\ra$; the ground state of the
two-electron system is a ``Wigner molecule'' in which the charge
density is strongly peaked near the vertices of the QD.
(b) The charge distribution of $|S_\updownarrow\ra$. 
As shown schematically in Fig.\ref{fig1}e, the electrons are\
localized in quadrants $a$ and $c$.
(c) The charge distribution of $|S_\leftrightarrow\ra$, shown
schematically in Fig.\ref{fig1}d.
As shown in \Eref{Sup_Sdown}, the ground state charge distribution (a)
is a superposition of (b) and (c).
(d) The charge distribution of $|S_1(V)\ra$ for $V=3\Delta_0$; for large values of the gating potential, the ground
state charge distribution strongly resembles the state $| S_\updownarrow \ra$.
In the same way $|S_2 (V) \ra$ correspondingly approaches 
$| S_\leftrightarrow \ra$ as $V$ is increased.}
\label{fig3}
\end{figure*}

For $V=0$ the charge density of the singlet ground state, $|S_1\ra$, for 
the square QD is shown in \Fref{fig3}(a). We see that
the charge density is distributed in sharp peaks, located near each
of the four corners. Perhaps surprisingly, 
the charge density for the excited singlet, $|S_2\ra$, has a practically
identical charge distribution.
This result may be understood by constructing (non-stationary) symmetric and antisymmetric superpositions of the eigenstates as
\begin{eqnarray}\label{Sup_Sdown}
  |S_\updownarrow\ra &=& \frac{|S_1(V=0)\ra-|S_2(V=0)\ra}{\sqrt{2}} \cr
  |S_\leftrightarrow\ra &=& \frac{|S_1(V=0)\ra+|S_2(V=0)\ra}{\sqrt{2}}
\end{eqnarray}
Note that throughout the paper the states $|S_\updownarrow\ra$ and $|S_\leftrightarrow\ra$ are both defined just for $V=0$ and so unlike $|S_1(V)\ra$ and $|S_2(V)\ra$ have no $V$-dependence. The charge densities of the states $|S_\updownarrow\ra$ and $|S_\leftrightarrow\ra$ are either localized near corners $a$ and $c$ or near corners $b$ and $d$, as shown in \Fref{fig3}(b) and \Fref{fig3}(c) respectively.
These states resemble the states shown schematically in \Fref{fig1}(c) and (d) which occur at finite $V$ when gate voltages are applied. The charge density for the state with $V=3\Delta_0$ is shown in \Fref{fig3}(d) and we see that it indeed looks very similar to the state $|S_\updownarrow\ra$. This similarity may be understood and quantified by expanding the eigenvectors $|S_1\ra$ and $|S_2\ra$ at finite $V$ as superpositions of states $|S_\updownarrow\ra$ and $|S_\leftrightarrow\ra$, defined for $V=0$. For a complete set of states at $V=0$ this expansion is of course exact, but since the lowest two singlets are well separated from higher excited states, we can expect truncation of the basis set to the 2D space of the lowest singlets to be a good approximation, provided $V$ is not too large. Within this 2D subspace the system may be described by the effective Hamiltonian:

\begin{equation}\label{Heff}
    H_S = E_{S_\updownarrow}(V)|S_\updownarrow\ra \la S_\updownarrow|+E_{S_\leftrightarrow}(V) |S_\leftrightarrow\ra \la S_\leftrightarrow|
    + \Delta (|S_\updownarrow\ra \la S_\leftrightarrow|+|S_\leftrightarrow\ra \la S_\updownarrow|).
\end{equation}
in which
\begin{eqnarray}\label{Energy_Heff}
    E_{S_\updownarrow}(V)&=& \la S_\updownarrow| H |S_\updownarrow\ra=E_{0S}+2V p^{bd}_{S_\updownarrow} \cr
    E_{S_\leftrightarrow}(V)&=& \la S_\leftrightarrow| H |S_\leftrightarrow\ra=E_{0S}+2V p^{bd}_{S_\leftrightarrow}
\end{eqnarray}
where $E_{0S}=E_{S_\updownarrow}(0)=E_{S_\leftrightarrow}(0)$ and
\begin{eqnarray}\label{parameter_Heff}
    p^{bd}_{S_\updownarrow} &=& \int_{bd} d\mathbf{r}_1 \int d\mathbf{r}_2  |\la \mathbf{r}_1, \mathbf{r}_2| S_\updownarrow \ra|^2, \cr
    p^{bd}_{S_\leftrightarrow} &=& \int_{bd} d\mathbf{r}_1 \int d\mathbf{r}_2  |\la \mathbf{r}_1, \mathbf{r}_2| S_\leftrightarrow \ra|^2, \cr
{ } &=& 1 - p^{bd}_{S_\updownarrow} .
\end{eqnarray}
The $bd$ on the integrations over $\mathbf{r}_1$ signifies restricting
the domain to quadrants $b$ and $d$, and the last step follows
from the square symmetry.
One can interpret $p^{bd}_{S_\updownarrow}$ as the probability that one electron is in quadrant $b$ or $d$, while the other electron is in any quadrant.
We expect this probability to be small in the strong correlation (large dot) regime since the amplitudes $\la \mathbf{r}_1, \mathbf{r}_2| S_\updownarrow \ra$ will all be small when $\mathbf{r}_1 \in \{ b,d\}$. Explicit calculations for the QD described in \Fref{fig2} give $p^{bd}_{S_\updownarrow}=0.109$. The tunneling terms $\Delta$ in the effective Hamiltonian \eref{Heff}, which rotate the configuration
between vertical and horizontal, can also be written as
\begin{equation}
    \Delta(V)=\la S_\leftrightarrow| H |S_\updownarrow\ra=\Delta_0+2V a_S
\end{equation}
where, $\Delta_0=\Delta(0)$, and again due to the square symmetry
\begin{equation}\label{Delta_Heff}
    a_S=\int_{bd} d\mathbf{r}_1 \int d\mathbf{r}_2  \la S_\updownarrow |\mathbf{r}_1, \mathbf{r}_2 \ra	\la \mathbf{r}_1, \mathbf{r}_2| S_\leftrightarrow \ra=0.
\end{equation}

Note that the effective Hamiltonian in \Eref{Heff} is a two-state tunneling Hamiltonian in which both electrons tunnel together with amplitude $\Delta=\Delta_0$, independent of $V$ in first order. When $V$ is large the lower energy state
simply corresponds to the two electrons mainly occupying quadrants $a$ and $c$,
 with a small probability $p^{bd}_{S_\updownarrow}$ of being in quadrants $b$ and $d$ where their potential energy is higher ($V$). Similarly, the higher energy state is when the two electrons mainly occupy quadrants $b$ and $d$,  with a small probability $p^{bd}_{S_\updownarrow}$ of being in quadrants $a$ and $c$ where their potential energy is zero.

Diagonalizing the effective Hamiltonian \eref{Heff} gives the eigenvalues
\begin{eqnarray}\label{spectrum_singlet}
    E_{S_1}&=&E_{0S}+V - \sqrt{[V(1-2p^{bd}_{S_\updownarrow})]^2+\Delta_0^2} ,  \cr
    E_{S_2}&=&E_{0S}+V + \sqrt{[V(1-2p^{bd}_{S_\updownarrow})]^2+\Delta_0^2} ,
\end{eqnarray}
with the corresponding eigenstates
\begin{eqnarray}\label{spectrum_eigenf}
    |S_1(V)\ra&=&+\cos(\theta)|S_\updownarrow\ra + \sin(\theta) |S_\leftrightarrow\ra, \cr
    |S_2(V)\ra&=&-\sin(\theta)|S_\updownarrow\ra + \cos(\theta) |S_\leftrightarrow\ra
\end{eqnarray}
where
\begin{equation}\label{theta_effe}
   \tan(\theta)=\frac{V(1-2p^{bd}_{S_\updownarrow})-\sqrt{[V(1-2p^{bd}_{S_\updownarrow})]^2+\Delta_0^2}}{\Delta_0}.
\end{equation}

We may analyze the $S_z=0$ triplets in a similar fashion though this is somewhat simpler, since the states $|T^0_\updownarrow\ra$ and $|T^0_\leftrightarrow\ra$ are not coupled by the Hamiltonian
\begin{equation}\label{Hamiltonian_triplet}
    H_T=E_{T_\updownarrow}(V)  |T^0_\updownarrow\ra \la T^0_\updownarrow | + E_{T_\leftrightarrow}(V) |T^0_\leftrightarrow\ra \la T^0_\leftrightarrow|.
\end{equation}
The eigenenergies are then
\begin{eqnarray}\label{spectrum_triplet}
    E_{T_\updownarrow}(V) &=&\la T^0_\updownarrow | H | T^0_\updownarrow \ra=E_{0T}+2V p^{bd}_{T_\updownarrow}, \cr
    E_{T_\leftrightarrow}(V) &=&\la T^0_\leftrightarrow | H | T^0_\leftrightarrow \ra=E_{0T}+2V (1-p^{bd}_{T_\updownarrow})
\end{eqnarray}
where
\begin{eqnarray}\label{p_t_triplet}
   E_{0T}&=&E_{T_\updownarrow}(0)=E_{T_\leftrightarrow}(0) \cr
\mbox{and} \quad p^{bd}_{T_\updownarrow}&=&\int_{bd} d\mathbf{r}_1 \int d\mathbf{r}_2 |\la \mathbf{r}_1,\mathbf{r}_2|T_\updownarrow\ra|^2.
\end{eqnarray}
For the QD referred to in \Fref{fig2} one obtains $p^{bd}_{T_\updownarrow}=0.142$.

\begin{figure}
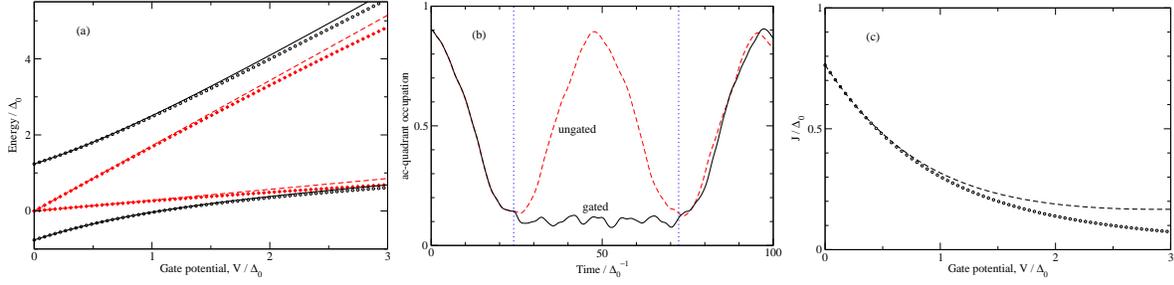

\begin{center}
\includegraphics[width=0.325\textwidth,clip=true]{fig4a}
\includegraphics[width=0.325\textwidth,clip=true]{fig4b}
\includegraphics[width=0.325\textwidth,clip=true]{fig4c}
\end{center}
\caption{(a) Comparison of the first-order perturbation theory
(\Eref{spectrum_singlet} and \Eref{spectrum_triplet}) with the exact
results. Perturbative results are shown with solid (black) / red (dashed)
lines for singlet / triplet states, while the exact results are given as
black circles / red diamonds. The agreement is excellent for
$ | V | < 2 \Delta_0$, while outside this range higher-order
corrections must be included.
(b) The QD is approximately prepared in the state $| S_\updownarrow \rangle$ by applying
$V = 3 \Delta_0$. This gate potential is then released and the singlet cycles
to a horizontal orientation and back again, as seen by the occupation
of the quadrants $ac$ (dashed line). If the gate potential is reapplied
during the cycle, the evolution of the singlet is frozen until the potential
is released again (solid line). The vertical (blue) dotted
lines indicate the times at which the gating potential is first 
applied and then removed. The ripples in the time-evolution result from
the excitation of energy levels outside the lowest multiplet.
(c) Exchange coupling $J/\Delta_0$ versus $V/\Delta_0$. Black circles show the exact
results, the dashed line the prediction from perturbation theory.}
\label{fig_comparison}
\end{figure}

To illustrate the accuracy of the simple effective Hamiltonian we compare it with the full numerical solutions of the 2-electron problem.
The energy eigenvalues vs $V$ for the lowest two singlets and triplets are plotted in \Fref{fig_comparison}(a) and we see excellent agreement between the effective model and the real one.  The advantage of the approximate model is that we have precise analytic solutions which can be used to derive analytic expressions for all quantities of interest.

\section{Singlet-triplet filtering \label{filter}}

The analytic expressions for the energies of the singlet states \eref{spectrum_singlet}
and the triplet states \eref{spectrum_triplet} give us a complete
picture for the system's time dependence. In particular 
they produce the phenomenon of ``singlet-triplet filtering'',
studied in detail in Ref.~\cite{bayat-QD} for the case of $V=0$.
This is a consequence of the very different dynamics displayed
by the singlet and triplet states under free evolution.
For example, if the system is initialized in the singlet state 
$| S_\updownarrow \ra$, it will subsequently
evolve coherently in time as
\begin{equation}\label{singlet-evolution}
e^{-iHt}|S_\updownarrow\ra = e^{-i(E_{0S}+V)t} \left( \sin(2\theta)
\sin(\omega t)  |S_\leftrightarrow\ra
    +  (\cos(\omega t)+i\cos(2\theta)\sin(\omega t))|S_\updownarrow\ra \right) ,
\end{equation}
where $\omega=\sqrt{\Delta_0^2+[V(1-2p^{bd}_{S_\updownarrow})]^2}$.
The horizontal and vertical components of the state  thus
cycle periodically in time, and for the specific case of
$V=0$ there will be a complete conversion of $|S_\updownarrow\ra$
to $|S_\leftrightarrow\ra$ after a time $t_R=\frac{\pi}{2\Delta_0}$.  
In contrast, if the system is initialized in the 
triplet state $|T_\updownarrow \ra$, its time dependence simply
consists of a trivial phase, as the triplet Hamiltonian \eref{Hamiltonian_triplet} does not contain
tunneling terms between the triplet states. The spin of the initial
state can thus be detected, or filtered, by a single charge measurement
at $b$ or $d$; the singlet component oscillates periodically with
time, while the triplet component stays frozen in position.

We now examine how the presence of the gate potential ($V \neq 0$) alters
this picture. When $V(1-2p^{bd}_{S_\updownarrow}) \gg\Delta_0$, 
$\theta \rightarrow 0$ and $|S_\updownarrow\ra$ becomes effectively the
eigenvector of the system and does not evolve. Applying a large
gate potential thus has the effect of shutting off the oscillation
of the singlet states between the vertical and horizontal configurations. 
We show this effect in \Fref{fig_comparison}(b), by plotting the time-evolution of the system prepared in the state $|S_\updownarrow\ra$ under
the full two-electron Hamiltonian \eref{H_schrodinger}. In the absence of
a gate potential, the singlet periodically cycles between its vertical 
and horizontal orientations as expected from our effective model. 
However, reapplying the gate potential freezes
the time evolution of the system, which remains halted until the potential
is again released.

\section{Qubits}

We define the two levels of our qubit as vertical singlet-triplet states, i.e. $|0\ra=|S_\updownarrow\ra$
and $|1\ra=|T_\updownarrow^0\ra$ (both $S_z=0$ states). In the regime of strong $V$ this qubit is well-defined, and is highly localized in its vertical configuration. 
For finite $V$, the eigenvector $|S_1\ra$ has a small contribution of $|S_\leftrightarrow\ra$, as $|\la S_\leftrightarrow|S_1\ra |^2\simeq |\frac{\Delta_0}{V(1-2p^{bd}_{S_\updownarrow})}|^2$, but this can be arbitrarily suppressed by controlling $V$. Furthermore, as shown in \Fref{fig2}, in the regime of strong $V$ both $|S_1\ra$ and $|T_1^0\ra$ become almost degenerate, and so there will be no relative phase between them.

\section{\label{single_qubit} Single qubit manipulations}

An arbitrary unitary operation on a single qubit can be realized by
sequential rotations around two different axes, such as $x$ and $z$.
Rotations around the $z$-axis may be simply achieved by the energy splitting $J=E_{0T}-E_{0S}$ between $|S_1\ra$ and $|T_\updownarrow^0\ra$ in the regime of $V(1-2p^{bd}_{S_\updownarrow})\gg \Delta_0$ where the electrons are still strongly localized in their vertical configurations (i.e. $|S_1\ra\approx |S_\updownarrow\ra$), but $J$ does not vanish. This exchange coupling $J$ generates a relative phase between the logical qubits $|0\ra$ and $|1\ra$ and thus performs a $z$-rotation.
In \Fref{fig_comparison}(c) the exchange coupling $J$ is plotted versus $V/\Delta_0$. From this figure
one can select the appropriate $V$ to give the $J$ that will perform the
desired rotation in a given time interval, during which the electrons
remain in the vertical configuration.

Rotation around the $x$ axis demands switching between $|0\ra$ and $|1\ra$. To do that a
gradient of magnetic field $\delta B_z$ is required between the vertical corners $ac$. There are
two different proposals for generating this gradient magnetic field: (i) polarizing the spin
of the nuclei in the bulk \cite{foletti}; (ii) using permanent micro-magnets \cite{pioro,forster}.
Here, we propose to use permanent micro-magnets near the the corners $ac$ as shown in \Fref{fig1}(a).
To perform an $x$ rotation, one has to push the electrons close to the micro-magnets to sense $\delta B_z$ by
applying a strong positive bias to the gates $G_a$ and $G_c$, which may be the micro-magnets themselves.
The gradient $\delta B_z$ rotates a single electron
around the $z$ axis and consequently switches between a singlet and a triplet state.

\section{Initialization}

An initial qubit state may be created by injecting a spin-up electron into corner $a$ and a down-spin electron
into corner $c$, while holding $V$ large enough to ensure that the electrons remain well-localized in these corners.
The electrons are thus created in the state $|+\rangle=(|0\ra+|1\ra)/\sqrt{2}$. Other initial states may then be generated with single-qubit
transformations, described earlier.

\begin{figure}
\centering
    \includegraphics[width=.75\textwidth,clip=true]{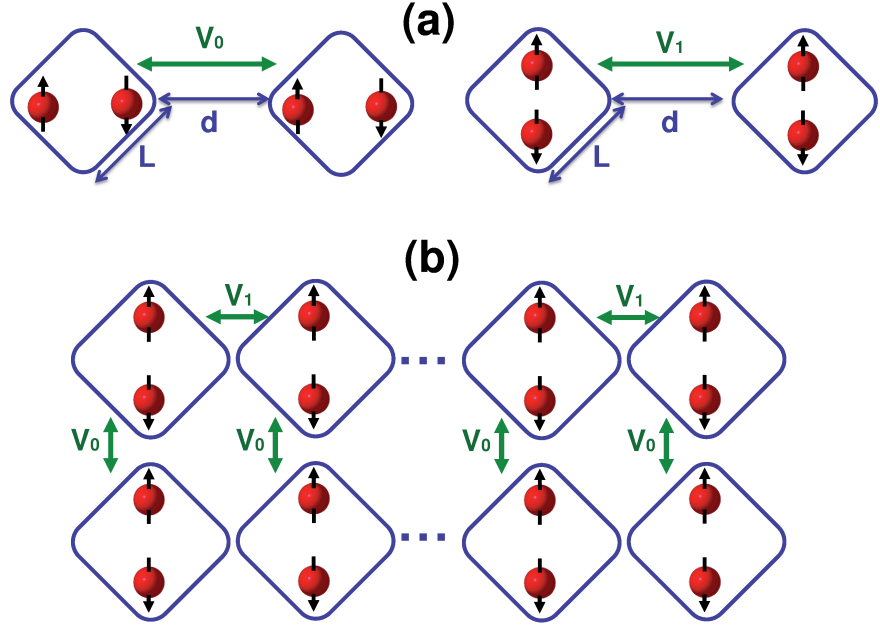}
    \caption{(a) Capacitive interaction of different charge configurations between two square quantum dots of size $L$ and distance $d$.
    (b) A two dimensional array of QDs for producing cluster states. Different horizontal and vertical interactions results in different two-qubit gates which can be compensated by local rotations.	 }
     \label{fig5}
\end{figure}

\section{Two-qubit entangling gate}

Apart from single qubit unitary operations, the
more challenging two-qubit quantum gates are also essential for universal quantum computation \cite{Bremner}.
We consider two square QDs, each containing two electrons encoding a singlet-triplet
qubit as described above. Interaction between the left and right qubits
is mediated through the electrostatic
Coulomb repulsion, as shown in \Fref{fig5}(a), which
is independent of the spin states.
Due to symmetry there are three independent electrostatic energies for the four
possible spatial configurations of electrons in two QDs (i.e. $\updownarrow\updownarrow$, $\updownarrow\leftrightarrow$, $\leftrightarrow\updownarrow$ and $\leftrightarrow\leftrightarrow$), where one of them can also set to be zero (overall energy shift). Therefore, as schematically shown in \Fref{fig5}(a), we can write the interaction between the two QDs as
\begin{equation}\label{Hint}
    H_I=\sum_{k,l=S,T} \left( u_0 |k_\leftrightarrow,l_\leftrightarrow\ra \la k_\leftrightarrow,l_\leftrightarrow|+
	u_1 |k_\updownarrow,l_\updownarrow\ra \la k_\updownarrow,l_\updownarrow| \right) ,
\end{equation}
where $u_0, u_1 >0$ account for the electrostatic Coulomb energies in the configurations  $\leftrightarrow \leftrightarrow$ and $\updownarrow \updownarrow$ respectively while the interactions of configurations $\updownarrow \leftrightarrow$ and $\leftrightarrow \updownarrow$ are chosen as the offset. By treating the electrons as classical point charges localized in the corners of the square confining potential, one can estimate $u_0$ and $u_1$ as functions of the dot size $L$ and their distance $d$ (see \Fref{fig5}(a)). In fact, the leading terms in Coulomb energies $u_0$ and $u_1$ are second order in $L/d$ giving
\begin{eqnarray}\label{U0_U1}
    u_0= \frac{e^2}{4\pi\epsilon_0 \epsilon_r d} \{ 3(\frac{L}{d})^2+ \mathcal{O}((L/d)^3) \} \cr
	u_1= \frac{e^2}{4\pi\epsilon_0 \epsilon_r d} \{ (\frac{L}{d})^2+ \mathcal{O}((L/d)^3) \},
\end{eqnarray}
where $\epsilon_0$ is the vacuum permittivity and $\epsilon_r=10.8$ is the dielectric constant of GaAs.

The Hamiltonian of the whole system then becomes $\Htot=H_L+H_R+H_I$, where $H_L$ and $H_R$ are given by \Eref{Heff} for the left and right QDs. The existence of $H_I$ changes the eigenstates of the system, and therefore the dynamics of \Eref{singlet-evolution}. To preserve the picture in which triplets do not evolve and singlet states rotate according to \Eref{singlet-evolution}, we should keep $u_0, u_1 \ll \Delta_0$ by fabricating the QDs relatively far apart. As discussed in Appendix A the choice of $u_0, u_1 \sim 0.1 \Delta_0$ is sufficient to retain this picture.

To have a two-qubit quantum gate, we first assume that $V$ is large and both qubits are initialized in an
arbitrary superposition of $|S_\updownarrow\ra$ and $|T_\updownarrow^0\ra$. To
operate the two-qubit gate, $V$ is set to zero. As $u_0,u_1\ll \Delta_0$ the interaction
Hamiltonian does not play an important role during this evolution, and so the dynamics is mainly
governed by $H_0=H_L+H_R$, in which the triplets do not evolve and singlets rotate according to
\Eref{singlet-evolution} with $V=0$. After time $t=t_R$ the evolution is again frozen by setting $V$ to a negative value which keeps the electrons in the horizontal configuration for an interaction time period of $t_I$, during which the system evolves under the action of $H_I$ alone. The potential barriers are then again removed (i.e. $V$ is set to zero) for
another period of $t=t_R$ to return the electrons to their initial positions.
One can write
the total evolution operator as
\begin{equation}\label{U_evolution_t}
    U(t_I)=e^{-iH_0t_R }e^{-iH_It_I}e^{-iH_0 t_R}.
\end{equation}
Over the interaction time $t_R<t<t_R+t_I$, each spatial configuration determined by the spin state of electrons has its own electrostatic energy, and thus the time evolution gives different phases to every state. One may easily verify that
\begin{eqnarray} \label{CZ-gate}
  U(t_I)|S_\updownarrow,S_\updownarrow\ra&=&e^{-iu_0t_I}|S_\updownarrow,S_\updownarrow\ra \cr
  U(t_I)|S_\updownarrow,T_\updownarrow\ra&=&- |S_\updownarrow,T_\updownarrow\ra \cr
  U(t_I)|T_\updownarrow,S_\updownarrow\ra&=&- |T_\updownarrow,S_\updownarrow\ra \cr
  U(t_I)|T_\updownarrow,T_\updownarrow\ra&=&e^{-iu_1t_I} |T_\updownarrow,T_\updownarrow\ra.
\end{eqnarray}
For $t_I=\frac{\pi}{u_0+u_1}$, this evolution realizes an entangling two-qubit gate such that its application to the state $|++\ra$ maximally
entangles the two qubits.
Moreover, this gate can be converted to the standard controlled $z$ ($CZ$) gate by two local rotations around the $z$ axis with the angle
$\frac{\pi}{2(u_0+u_1)}$.

\section{Readout}

In our mechanism, single qubit measurement in the computational $z$ basis is
the singlet-triplet
measurement of the electron pair in the QD. This can be achieved
by setting $V$ to zero, thereby allowing tunneling from
vertical to horizontal configurations for the singlet
(triplet states are unable to tunnel from vertical to horizontal
as there are no tunneling elements between these states).
A single charge detection then fulfills the singlet-triplet measurement~\cite{bayat-QD}, as explained in Section \ref{filter}.
Single qubit measurement in any other basis can be simply reduced to a $z$ measurement by applying proper local rotations.

\section{Applications}

Universal quantum computation can be achieved in two dimensional network of qubits, which can be prepared
in a highly entangled state termed a cluster state \cite{cluster-state}. To prepare a
cluster state we need a  two-dimensional array of qubits all initially prepared in
$|+\ra$ states. Then a homogeneous action of $CZ$ gates between all neighboring qubits
generates a cluster state, on which measurement-based quantum computation
can be realized by local rotations and single qubit measurements \cite{cluster-state}.
Such an array of QDs is shown in \Fref{fig5}(b).
Note that when electrons are frozen in their locations, the electrostatic interactions
only give a global phase. In this structure when the
system is released for a global gate operation, the type of the gate that acts on rows is
different from the one acting on columns, unless $u_0=u_1$.
However, these gates can be locally transformed to $CZ$ gates, and thus
the outcome is still a cluster state and can be used for measurement-based
quantum computation. Note that in an array of double dots used to realize a two-qubit gate, one has to change the charge configurations to $(0,2)$, which makes the left and right neighbor qubits experience asymmetric interactions, thereby prohibiting simultaneous identical gates.

\section{\label{practicality} Practicality and time scales}

In this section we estimate the parameters of the system and explore the 
experimental feasibility of our proposal. For QDs of side-length
$L=400$ nm, we have $\Delta_0 \simeq 20 \mu$eV. Separating the QDs so
that $u_0+u_1 \simeq 2 \mu$eV guarantees the validity of \Eref{singlet-evolution} to very high precision, as  $\Delta_0/(u_0+u_1) \simeq 10$. The operation time of our two-qubit gate is $2t_R+t_I=\frac{\pi\hbar}{\Delta_0}+\frac{\pi\hbar}{u_0+u_1}$. Using the above values yields an
operation time of $2t_R+t_I=12$ ns. The spacing between
the QDs ($d$) corresponding to this choice of physical parameters can
be calculated from \Eref{U0_U1}, giving a value of $d\simeq 3.6 \mu$m.

To determine the temperatures in which the system can operate one has to estimate the energy gap of the system. From \Fref{fig2} one can see that the energy gap between the singlet and triplet subspace is $\Delta E\simeq \Delta_0$. For the system to operate safely the temperature should be below the energy gap. Using the above parameters for a square dot of size $L=400$ nm in which $\Delta_0 \simeq 20 \mu$eV, one can evaluate the energy gap as $\Delta E\simeq 200$ mK which is larger than the typical temperatures ($\sim 100$ mK) achievable with current dilution fridges. This clearly shows that the proposed mechanism can be realized with existing technology.

\section{Decoherence and robustness}

 A major obstacle for realizing two qubit gates through capacitive interaction in double dot systems is the very short charge dephasing time ($\sim 1$ ns).
Such a short time scale is due to the interaction between the electric dipole $\overrightarrow{\mathbf{p}}$ of the two electrons with the random fluctuating electric field $\overrightarrow{\mathbf{E}}(t)$ (namely	$U=-\overrightarrow{\mathbf{p}}.\overrightarrow{\mathbf{E}}(t)$).
To realize a two-qubit quantum gate in double dot systems one can use the electrostatic coupling between the two neighboring double dots which give different phases to singlets and triplets according to their different charge configurations. Since the charge configurations of the singlets and triplets are quite similar for the $(1,1)$ configuration the time scale of the two qubit gate becomes too long (for instance it is $\sim 150$ ns in the realization of Ref.~\cite{yacobi-2QBgate}). One can speed up this process significantly, even up to $\sim 20$ ns \cite{marcus-CZ}, by giving more offset energy to one of the dots in order to convert the singlet charge configuration to $(2,0)$, leaving one of the dots empty, to make the capacitive interaction stronger. However, this
produces significant charge dephasing, as singlets and triplets have different electric dipole moments (due to the asymmetric charge configuration $(0,2)$ of the singlets) and interact differently with electric field fluctuations. In contrast, in our square QD proposal the charge configuration always remains symmetric with zero electric dipole moment. Hence, the leading term for charge dephasing is quadrupolar, giving a much longer charge dephasing time.

The hyperfine interaction between the electrons and nuclei in the bulk is the main source of decoherence in QDs.
To compensate this effect we may use the recently-implemented idea of multiple-pulse echo sequence \cite{Yacoby-coherence}.
In this technique the quantum states of the two electrons are swapped through exchange interaction regularly, allowing decoherence times of  $T_2\sim 260 \mu$s.
As mentioned in the previous section, for dots with the size $L=400$ nm, we have $\Delta_0 \simeq 20 \mu$eV and fabricating the dots with a spacing of $d\simeq 3.5 \mu$m gives $\Delta_0/(u_0+u_1) \simeq 10$. These parameters imply that the two-qubit operation time will be $2t_R+t_I\simeq 12$ ns which allows for of the order of $10^5$ operations within the coherence time of the system. Even in the absence of regular exchange of quantum states, the hyperfine interaction between the electrons and nuclei in the bulk is at least two orders of magnitude smaller than $\Delta_0$ \cite{bayat-QD}, and one order of magnitude less than $u_0+u_1$. This
guarantees that it has no significant effect over the proposed fast dynamics ($\sim$ 12 ns) although the coherence time is then limited to 1 $\mu s$ \cite{petta} and thus the number of operation reduces to $\sim 10^3$.

Another major of imperfection in singlet-triplet double QD systems is due to the gate voltage fluctuations \cite{Das-Sarma}. This makes the tunneling between the two dots noisy which then results in fluctuations in spin exchange coupling which is $J\sim t^2/U$ (for tunneling $t$ and on-site energy $U$). As tunneling in double dot systems is directly controlled by gate voltages while the on-site energy is independently determined by the Coulomb interaction, the spin exchange coupling fluctuates in time with the tunneling \cite{Das-Sarma}. In our square QD system, however, the exchange coupling $J=E_{T_\updownarrow}(V)-E_{S_1}(V)$ is determined by $J\sim \Delta_0^2/V$. So, in the large $V$ limit, which we use for the single qubit gate operations, the fluctuations of the gate voltage $V$ appear only in the denominator and $\Delta_0$ is independent of $V$ to first order. Thus we expect less sensitivity to gate voltages in our scheme.

So far we have assumed that $V$ can be instantaneously switched on and off at desired times. In reality gate voltages cannot jump instantly, and so $V$ varies gradually. One can estimate the gradual switching error by assuming that $V$ is switched off (or on) linearly over a period of $\tau$. For instance, in \Eref{singlet-evolution} a linear switching of $V$ over the time period $t=t_R$ to $t=t_R+\tau$ produces an error equal to
$\sin^2(\frac{\Delta_0 \tau}{2})\approx \frac{\Delta_0^2 \tau^2}{4}$. In particular, for QDs of size $L=400$ nm, (i.e. $\Delta_0=20 \mu$eV) a gradual switching with duration $\tau=10$ ps induces less than $2\%$ error in our desired state.

\section{Alternative realization}

Apart from GaAs technology, one can also realize our quantum cellular automata using the silicon atom dangling bonds on hydrogen terminated silicon crystal surface \cite{si-dangling-bond,Wolkow-Si-DB}. The four coupled QDs located in a ring, hosting two highly interacting electrons (fully capable for achieving our spin filtering dynamics) have been realized experimentally \cite{Wolkow-Si-DB}. The isotopically purified silicon provides very long decoherence time ($T_2$ exceeding 200 $\mu s$) as
the nuclear spin interaction is practically eliminated, and a charge dephasing time of $\sim 200$ns has been measured for charge qubits in Si double QDs \cite{williams}.

\section{Conclusions}

We have shown that the singlet and triplet states of a pair of electrons held in a
square QD can be used as a rapid and deterministically-controlled qubit.
Introducing electrostatic interactions between neighboring qubits allows two-qubit
entangling gates to be constructed, thus enabling universal quantum computation, with particular suitability to its measurement-based
version. The extra charge orbital in our system enables the fundamental issue of short charge dephasing time of singlet-triple qubits in double dots to be tackled by having zero dipole moment. Hence the leading interaction with the environment is quadrupolar, resulting in much longer charge coherence.
The architecture was inspired by classical cellular automata implementations \cite{lent-porod,Cowburn} thereby linking them to the quantum realm and providing a path for quantum-classical integrability in computer technology. While singlet-triplet qubits have already been realized in double dots, our proposal makes a number of advances, namely: (i) no electric dipoles at any stage (potentially much longer coherence); (ii) symmetric gate operations; (iii) no relative phases during the storage; and (iv) non-adiabatic (i.e. fast) operation which could become preferable in view of the continually improving speed of control electronics; (v) less sensitivity to gate voltage fluctuations. We should emphasize that 
as well as the single QD geometry we have presented here, our results
are also applicable with minor modifications to the four-dot structure 
studied experimentally in  Ref. \cite{Charles-fourDot}.

\smallskip

\ack AB was supported by the EPSRC grant EP/K004077/1 (nano-electronic based quantum technologies). SB is supported by an ERC grant. CEC was supported by the MINECO (Spain) through grants FIS2010-21372 and FIS2013-41716-P. MP thanks the EPSRC, and JHJ and CEC acknowledge support from the EU NanoCTM network.

\appendix

\section{Two neighboring QDs} \label{appendix2}

In the paper we have introduced the interaction between two neighboring QDs which interact through capacitive Coulomb repulsion.
The form of the interaction Hamiltonian is given by $H_I$ in \Eref{Hint}. The existence of $H_I$ changes the eigenvectors of the system and therefore may affect the system's dynamics.
To quantify the effect of this interaction on the spectrum of $H_0=H_L+H_R$, we compute the modified eigenstates of the whole system $H_{tot}$ when $H_I$ is treated perturbatively. Within this regime the new  relevant unnormalized eigenvectors are given to first order by
\begin{eqnarray}\label{eigenvecs}
    |S_1,S_1\ra &\rightarrow&  |S_1,S_1\ra + a |S_1,S_2\ra + a |S_2,S_1\ra - b |S_2,S_2\ra \cr
    |S_1,S_2\ra &\rightarrow&  \frac{|S_1,S_2\ra + |S_2,S_1\ra}{\sqrt{2}} - a\sqrt{2} (|S_1,S_1\ra - |S_2,S_2\ra) \cr
    |S_2,S_1\ra &\rightarrow&  \frac{|S_1,S_2\ra - |S_2,S_1\ra}{\sqrt{2}} \cr
    |S_2,S_2\ra &\rightarrow&  |S_2,S_2\ra+ b |S_1,S_1\ra - a |S_1,S_2\ra - a |S_2,S_1\ra  \cr
    |S_1,T_0\ra &\rightarrow&	|S_1,T_\updownarrow^0\ra - c  |S_2, T_\updownarrow^0\ra  \cr
    |T_0,S_1\ra &\rightarrow&	|T_\updownarrow^0,S_1\ra - c  |T_\updownarrow^0 ,S_2\ra  \cr
    |S_2,T_0\ra &\rightarrow&	|S_2,T_\updownarrow^0\ra + c  |S_1, T_\updownarrow^0\ra  \cr
    |T_0,S_2\ra &\rightarrow&	|T_\updownarrow^0,S_2\ra + c  |T_\updownarrow^0 ,S_1\ra
\end{eqnarray}
where,
\begin{eqnarray}\label{parameters}
    a = \frac{u_0-u_1}{8\Delta_0}, \ \ b =  \frac{u_0+u_1}{16\Delta_0}, \ \ c=\frac{u_1}{4\Delta_0}.
\end{eqnarray}

Tuning $u_0,u_1\simeq 0.1 \Delta_0$ guarantees that the oscillation between $|S_\updownarrow\ra$ and $|S_\leftrightarrow\ra$ remains valid up to a very high fidelity ($>0.9$).

\end{document}